\documentclass[twocolumn,superscriptaddress,amssymb,amsmath,aps,prl]{revtex4-2}

\usepackage{color}
\usepackage{comment}
\usepackage{graphicx}
\usepackage{tensor}

\newcommand{\Mp}{m_{\text{\tiny P}}}
\newcommand{\rhop}{\rho_{\text{\tiny P}}}

\begin{document} 

\title{ Trace anomaly and compact stars. }

\author{Ignacio A. Reyes}
\affiliation{Institute for Theoretical Physics, University of Amsterdam, PO Box 94485, 1090 GL Amsterdam, The Netherlands}

\begin{abstract}

A widespread assumption states that quantum effects of matter in curved spacetimes become relevant only when the inverse curvature radius approaches the Planck mass $\Mp$. We challenge this view by showing that, relying solely on universal properties of QFTs at high energies, the trace anomaly becomes macroscopic instead at energy scales $ \left( m/\Mp \right)^{\frac{1-\alpha}{2-\alpha}}  \Mp$, where $m\ll \Mp$ and $\alpha<1$ are, respectively, the mass scale and exponent appearing in the matter equation of state at high energies. As an application, we consider compact stars close to their Buchdahl limit, and examine the propagation of scalar waves on the resulting backreacted geometries. At the aforementioned energy scale, the curvature becomes negative within a neighborhood of the the star's center, implying a qualitative change in the behavior of the associated spectrum of modes.

 \end{abstract}

\maketitle

\section{Introduction }

Understanding the relationship between General Relativity and Quantum Field Theory (QFT) remains one of the most intriguing questions in theoretical physics. A beautiful example of the interplay between quantum theory and geometry is the subject of degenerate stars which is grounded in quantum statistical mechanics and the Pauli exclusion principle. %The Chandrasekhar and Tolman-Oppenheimer-Volkoff limits for the maximum mass of white dwarfs and neutron stars respectively use relativistic equations of state for fermions. 

A commonly held view is that above the neutron density, there is little we can say about the behavior of relativistic stars because we ignore the details of the QCD equation of state at such great pressures and densities. Although partially correct, this view overlooks two essential points. The first is that quantum statistical mechanics fails to reproduce the existence of \textit{anomalies}, a vital component of our modern understanding of QFT. The second is that, at energies much higher than the particles' mass, the equation of state simplifies because the theory approximates a Conformal Field Theory (CFT). 

In this context it is frequently argued that QFT effects such as vacuum polarization can only become macroscopic at the Planck scale. This is a misconception. As we will see, the fact that matter at very high energies becomes conformal at leading order, implies that the anomalous contributions to the curvature scalar become macroscopic at energies much below the Planck mass. 

%The celebrated Hawking effect, describing the evaporation of a black hole, is too small to be detected for stellar mass objects. In this work we describe a closely related effect, which instead takes place in the interior of regular compact stars.   

We will now review these ingredients in order to incorporate them into the semi-classical analysis. 

%%%%%%%%%%%%%%%%%%%%%%%%%%%%%%%%%%%%%%%%%%%%%%%%%%%%%

\section{Energy-momentum in QFT}\label{secQFT}

Anomalies describes in detail how certain `classical' symmetries fail after quantization, and are intimately connected to topology. They appear when quantum fields interact with gauge fields. Since General Relativity is a gauge theory, quantum fields in curved spacetimes develop anomalies. 

In this paper we will focus on the dynamics of the \textit{trace (or Weyl) anomaly}. For QFTs propagating in curved spacetime, the total stress tensor can be decomposed as
\begin{align}\label{TTT}
T_{\mu\nu}=T_{\mu\nu}^{(m)}+T_{\mu\nu}^{(A)}\,.
\end{align}
We shall omit the expectation value and normal ordering, as they are to be understood everywhere, i.e. $\langle :T_{\mu\nu}:\rangle=T_{\mu\nu}$, and specify the choice of quantum state below. 

Here  $T_{\mu\nu}^{(m)}$ corresponds to the terms that break conformal symmetry already at the classical level and do not explicitly depend on the metric, e.g. a mass term (hence the $m$). Although ultimately all matter in the universe is quantum, we shall loosely refer to this term as `classical'. On the other hand, $T_{\mu\nu}^{(A)}$ stands for the anomalous contribution, and depends explicitly on curvature. Its components are mostly unknown except in highly symmetric geometries. We will refer to these as the `quantum' contributions.

For example, in the theory of relativistic stars, the `classical' piece $T^{(m)}_{\mu\nu}$ is simply the contribution of the fluid arising from the Fermi pressure, and the background source that generates the Oppenheimer-Volkoff metric\,\cite{Oppenheimer:1939ne}. Although this term originates in the exclusion principle and cannot be explained without quantum theory, once we assume the Fermi-Dirac distribution it can be treated otherwise as a classical fluid. In contrast, $T_{\mu\nu}^{(A)}$ involves the more technical procedure of renormalizing a QFT on the background given by the O-V solution.

%The separation done in \eqref{TTT} does not correspond to `classical plus quantum'. All matter in the universe is quantum. The distinction is between terms that break conformal symmetry already at the classical level, and those that respect it classically but break it after quantization -- the anomalous terms.... which are the more involved components that arise from regularization QFTs on curved space.

%For example, if a star is made of degenerate fermions (see below) the Fermi pressure -- which is quantum and depends on the fermion's mass -- is contained in $T^{(m)}$, not in $T^{(A)}$. 

In $3+1$ dimensions the conformal or trace anomaly states that, in the trace of the full stress tensor
\begin{align}\label{Ttrace}
\tensor{T}{^\mu_\mu}\equiv T^{(m)}+T^{(A)}\,,
\end{align}
in addition to the `classical' term $T^{(m)}$, the renormalized contribution is given by\,\cite{Duff:2020dqb}
 \begin{align}\label{Tmumu}
   T^{(A)}= \frac{1}{(4\pi)^2}\left( c W^2 - a E  \right)\,.
\end{align}
Here $W^2$ is the square of the Weyl tensor, and $E=R^{\mu\nu\rho\sigma}R_{\mu\nu\rho\sigma}-4R_{\mu\nu} R^{\mu\nu}+R^2$ is the Euler class, which integrated on the entire manifold yields the Euler characteristic $\chi$. The two coefficients $c,a>0$ are well understood and characterize the particular theory in question. We have absorbed their $\hbar$ dependence into the gravitational constant to form the Planck mass as shown below, so that $a,c$ are dimensionless and roughly order unity.
 
We have not included a term $\Box R$ since its coefficient is renormalization-scheme dependent, and furthermore its presence would imply that the trace of the field equations would be higher than second order. $c$ will play no major role here. The numerical value of $a$ will not be important, but the fact that it is positive is crucial. In \eqref{Tmumu} we have omitted the extra contribution due to the reference quantum state, because this is already accounted for in $T^{(m)}$. %Finally, the anomalous stress tensor is identically conserved, $\nabla_\mu \langle T^{(A)\mu\nu}\rangle=0$, because it is a purely geometric object derived from a diffeomorphism-invariant effective action\,. Therefore the continuity equation still reads $\nabla_\mu T^{(m)\mu\nu}=0$.

Having introduced the basic framework -- the trace anomaly of QFTs in curved spacetime -- we are ready to specify the assumptions we will use in our derivation below.

\subsection{Assumptions}
For simplicity we consider matter composed of an isotropic perfect fluid  
\begin{align}\label{Tmmunu}
\tensor{T}{^{(m)}^\mu_\nu}=\text{diag}(-\rho,p,p,p)\,.
\end{align}
We assume that much below the Planck scale $\rho\ll \Mp^4$,   

\begin{enumerate}

\item[] (A1) QFT in curved spacetime, via the semi-classical equations 
\begin{align}\label{EEna}
G_{\mu\nu}=8\pi G \left( T^{(m)}_{\mu\nu} + T^{(A)}_{\mu\nu} \right)\,,
\end{align}
provides the correct physical description of the system, with the geometry regarded as classical. 

\item[] (A2) At energies much higher than the mass of the matter degrees of freedom, the matter equation of state behaves as
\begin{align}\label{Tq}
T^{(m)} \underset{ m^4\ll \rho \ll \Mp^4}{\approx } -m^{4(1-\alpha)} \rho^{\alpha}
\end{align}
where $\alpha<1$, and $m\ll \Mp$ is a mass scale appearing for dimensional reasons.

\item[] (A3) The individual components of the anomalous contributions are subleading compared to the non-anomalous ones,
\begin{align}\label{A<M}
|T^{(A)}_{\mu\nu}|\ll |T^{(m)}_{\mu\nu}|\,.
\end{align}

\end{enumerate}

Let us now comment on these assumptions. (A1) is self-evident. We will only make statements about physics much below the Planck scale.

(A2) describes the equation of state at very high energies, and ensures that matter behaves like a CFT at leading order. The corrections depend on the mass $m$ of the fields involved, which for all known elementary particles is minuscule compared to $\Mp$. Two influential examples that \textit{do not} satisfy this condition are the Oppenheimer-Snyder collapse model (with $p=0$ but $\rho>0$), and Schwarzschild's constant density star. On the other hand, as we will see in more detail below, an important example that \textit{does} satisfy \eqref{Tq} is the Fermi gas.

Regarding (A3), there are almost no known results for $T^{(A)}_{\mu\nu}$ in $3+1$ dimensions, as the computations involved are plaged with renormalization ambiguities. For the known case of conformally coupled fields in a conformally flat background\,\cite{Brown:1977sj}, the renormalization scheme-independent terms are quadratic in the curvature tensors. Thus, a (naive) direct comparison of the individual components shows that they could only become of the same order in \eqref{A<M} at the Planck scale. Moreover it implies that the semi-classical equations \eqref{EEna} will remain effectively second order. 

(A3) means that the leading order contribution to the individual components of $g_{\mu\nu}$, $G_{\mu\nu}$, etc. is determined by $T^{(m)}_{\mu\nu}$ in \eqref{EEna}. However, (A3) by no means implies that all \textit{combinations} of the components of the respective stress tensors satisfy an inequality like \eqref{A<M}. Indeed, due to (A2), the \textit{trace} of the stress tensor clearly doesn't. %Take the case of isotropic perfect fluids as an illustration. (A3) implies that $|\rho^{(m)}|\gg |\rho^{(A)}|$ and $|p^{(m)}|\gg |p^{(A)}|$. But this does not imply a hierarchy between $T^{(m)}=-\rho^{(m)}+3p^{(m)}$ and $T^{(A)}=-\rho^{(A)}+3p^{(A)}$, because due to (A2), the former trace vanishes at leading order whereas the latter doesn't.   

Therefore we will focus on the trace of Einstein's equations,
\begin{align}\label{quantumtrace}
R=-8\pi G \left( T^{(m)}+T^{(A)} \right)\,.
\end{align}
Notice that in the absence of the anomaly, $R>0$ by assumption of the equation of state in \eqref{Tq} ($T^{(m)}<0$). We emphasize again that by (A3) we are \textit{not} assuming any dominance between the traces $T^{(m)}$ and $T^{(A)}$. In the next section we evaluate the anomalous contributions in certain circumstances of physical interest.

\section{Anomalous Einstein equations.}

In order to assess the effect of the anomalous trace in the field equation \eqref{quantumtrace}, let us rewrite it in a more convenient form. Using the Ricci decomposition, the Euler density is given by
\begin{align}\label{E1}
E&=W^2+2\left( \frac{R^2}{3}-R_{\mu\nu} R^{\mu\nu} \right)\,.
\end{align}
Here, the curvature tensors correspond to the metric that solves \eqref{EEna}. Although in general evaluating \eqref{E1} is a formidable task, we will make use of two important simplifications.

First, we will restrict our focus to cases where $W^2\approx 0$. This will be either exact or a good approximation in the cases of interest. If the metric is conformally flat, such as Friedmann-Robertson-Walker or the Schwarzschild constant-density star, the Weyl tensor vanishes identically. Moreover, for a regular spherically symmetric static spacetime, $W^2=0$ at the origin (see comments after \eqref{conditions}). So by focusing only on what happens in a vicinity of the origin, we can neglect the Weyl contribution.

The backreacted semi-classical equations then read
\begin{align}\label{eqR}
R&=-8\pi G \left( T^{(m)}-aE \right)\,.
\end{align}
Now clearly the $R^2$ term appearing in \eqref{E1} will not correct the LHS of \eqref{eqR} before the Planck scale, so we can ignore it. On the other hand, we notice that
\begin{align}\label{}
R_{\mu\nu}R^{\mu\nu} \overset{(A3)}{\approx} (8\pi G)^2  T^{(m)}_{\mu\nu} T^{(m)\mu\nu} \overset{(A2)}{\sim} G^2\rho^2\,.
\end{align}
Here in the first step we have used (A3), and then from (A2) we used that at leading order $p\sim \rho$. Therefore at leading order in density the Euler class becomes
\begin{align}\label{Eulercenter}
E&\sim -G^2 \rho^2\ \ \ \ \ \ \ \rho\gg m^4\,.
\end{align}

The two essential features of the Euler density \eqref{Eulercenter} are that: 1) it is negative definite and 2) it grows quadratically with the density. \textit{None of these properties can be varied in any way}: they are completely fixed by the internal consistency of QFT. These features will prove essential below.

Now that we have evaluated $E$, we are in position to examine its backreaction on the geometry. The trace equation \eqref{quantumtrace} becomes, to lowest orders in $\rho$
\begin{align}\label{R0q}
R\approx \Mp^{-2} \left( m^{4(1-\alpha)} \rho^\alpha - \Mp^{-4} \rho^2  \right)%\approx \kappa \left( C \rho^\alpha - \frac{a}{(4\pi)^2}\kappa^2 \rho^2  \right)
\end{align} 
where we have introduced the Planck mass $\Mp=1/\sqrt{G}$, set $a\sim 1$ and again dropped prefactors of order unity, as we are only interested in scalings and orders of magnitude.

From \eqref{R0q} we can see the anomaly at work: At low densities, the scalar curvature is positive, consistent with the non-anomalous equation of state (A2). It grows sub-linearly, reaches a maximum,  and then becomes negative due to the anomaly. The crossover density where $R=0$ is given by 
\begin{align}\label{rhoc}
\rho_c\sim   \left( m/\Mp \right)^{\frac{4(1-\alpha)}{2-\alpha}} \rhop
%\sim \left( \frac{C}{\kappa^2} \right)^{\frac{1}{2-\alpha}} \ll \rhop=\Mp
\end{align}
or equivalently in terms of energy,
\begin{align}\label{}
E_c\sim  \left( m/\Mp \right)^{\frac{1-\alpha}{2-\alpha}} \Mp\,.
\end{align}

This is the main result of this paper. It applies whenever the Weyl squared is negligible, e.g. to both cosmology and the interior of compact static stars. Given an equation of state specified by $m$ and $\alpha$, \eqref{rhoc} sets the regime of density/pressure where QFT in curved spacetime, not merely relativistic quantum statistical mechanics, becomes relevant. Although $\rho_c$ can be enormous, it is still well below the regime where quantum gravity is necessary because $\alpha<1$ and $m\ll \Mp$. Ultimately, this suppression is originated in \eqref{quantumtrace}: in the high energy limit, $T^{(m)}$ scales \textit{slower} than $\rho$ because it must vanish at leading order (we approach a CFT fixed point), whereas $T^{(A)}$ scales quadratically. 

In summary, the answer to: ``When do individual components $T^{(m)}_{\mu\nu}$ and $T^{(A)}_{\mu\nu}$ become the same order?" is generically the Planck density. However, if we ask instead ``When do the \textit{traces} of these stress tensors are of the same magnitude?", the answer is \eqref{rhoc}, much before the Planck scale. The problem then becomes identifying physical phenomena that are sensitive to the curvature scalar only. We will now focus on a concrete physical context where this takes place: the interior of very compact relativistic stars.  

%%%%%%%%%%%%%%%%%%%%%%%%%%%%%%%%%%%%%%%%%%%%%%%%%%%%%%

\section{The Generalized Buchdahl Limit}\label{sec:GBL}

The current paradigm of gravitational collapse is based on the Oppenheimer-Snyder model\,\cite{Oppenheimer:1939ue} consisting of a pressureless cloud of dust in free fall. On the opposite side of dynamics, one can consider slow adiabatic contraction\,\cite{Bondi:1964zza}. The endpoint of this process is \textit{not} a black hole, but the (generalized) Buchdahl limit characterized by central curvatures growing without bound. We wish to explore the consequences of the anomaly described above in these backgrounds.

Consider a sphere of mass $M$ and coordinate radius $r_b$, modeled again as an isotropic perfect fluid, with stress tensor \eqref{Tmmunu} on a static, spherically symmetric geometry
\begin{align}\label{ds2static}
   ds^2 &=-f(r) dt^2+h(r)dr^2+r^2 d\Omega_2^2\,,
\end{align}
subject to the regularity conditions at the origin
\begin{align}\label{conditions}
h(0)=1\ ,\ f'(0)=0=h'(0)\,.
\end{align}
These conditions imply that $W^2=0$ at $r=0$. The geometry outside the sphere is the Schwarzschild vacuum.

Buchdahl's theorem is a beautiful application of GR coupled to static matter\,\cite{Buchdahl:1959zz}. Assuming only that $\rho>0$, $\partial_r\rho\le0$ and that Einstein's equations hold, the requirement that the metric is everywhere regular (i.e. $f(r)>0$) leads to
\begin{align}\label{BL}
\frac{r_b}{GM}\ge \frac{9}{4}\,.
\end{align}

Usually, the saturation of the bound is known as the Buchdahl limit. We \textit{define} the Generalized Buchdahl Limit (GBL) as follows: given any equation of state, the GBL is given by the limiting solution with $f(0)\to 0^+$ of infinite central redshift. It corresponds to the `last smooth' (or first singular) static solution to Einstein's equations.  
 
We have defined it in this way in order to distinguish it from the more common notion of Buchdahl limit, which is the saturation of \eqref{BL}, for the following reason. For any given equation of state, the limit $f(0)\to 0^+$ does not necessarily saturate the Buchdahl bound $\frac{r_b}{GM}=\frac{9}{4}$. Buchdahl's theorem only tells us that under the given assumptions, the ratio $r_b/(GM)$ at the corresponding GBL will be larger or equal than $9/4$. 

For example, for uniform density stars the GBL coincides with the standard Buchdahl Limit, while for Tolman type IV,V,VI metrics the GBL has $r_b/(GM)=3,4,\frac{14}{3}$ respectively\,\cite{Tolman:1939jz}. This happens of course because all these solutions have different equations of state. Also, the Buchdahl limit (and thus the GBL) happens well before reaching the Schwarzschild radius $r_b=2GM$, so black holes are not relevant in this context. 

As the star becomes more compact, the central density and pressure will increase without bound as we approach the GBL. To see this, it suffices to examine the `classical' conservation equation $\nabla_\mu T^{(m)\mu\nu}=0$ (recall that the `quantum' contribution is identically conserved\,\cite{Wald:1984rg}). Integrating $\nu=r$ (the only non-vanishing component) throughout the entire sphere yields
\begin{align}\label{logff}
\log \frac{f(r_b)}{f(0)}=-2\int_{p(r_b)}^{p(0)} \frac{dp}{p+\rho}\,.
\end{align}

In general the equation of state of matter will take the form of a polytrope $p\sim \rho^\gamma$ with $\gamma\geq 1$ varying throughout the star, with $\gamma$ closer to unity  as matter approaches the conformal limit. The contributions coming from the outer layers of the star associated to $p(r_b)=0$ thus remain finite. In the GBL, $f(r_b)$ remains finite (due to Buchdahl's theorem), while $f(0)\to 0$ by definition of the GBL. Therefore from \eqref{logff} we conclude that $\rho(0),p(0)\to \infty$ in this limit. Thus, the GBL is a natural scenario involving very high energies where the anomalous QFT effects discussed earlier can play an important role. 

Now that we have established the behavior of the fluid in the GBL in general, let us focus on a very important example, degenerate matter, which is essential in our explanation of compact stars.

\subsection{Example: cold Fermi gas}

The equation of state of a non-interacting gas of fermions can be derived using statistical mechanics, the stress tensor taking the form \eqref{Tmmunu}. At low energy density 
\begin{align}\label{low}
p\propto \rho^{5/3}\ \ \ \rho\ll m ^4
\end{align}
while at densities much higher than the fermion's mass,
\begin{align}\label{ultra}
%\rho-3p\approx m^2 \sqrt{\rho}+\hdots\ \ \ \rho\gg m^4
3p= \rho-m^2 \sqrt{\rho}\ \ \ \rho\gg m ^4
\end{align}
at next-to-leading order. This corresponds to $\alpha=1/2$ in \eqref{Tq}. 

A comment about the equation of state is in order. In the literature it often stated that the `ultra-relativistic' regime satisfies $p\sim \rho^{4/3}$. However, this is rather misleading. This result is obtained by keeping the exact relativistic expression for the pressure, but approximating the energy density by the mass contribution only. This is an effective description valid at intermediate energies, while \eqref{ultra} is the truly ultra-relativistic regime. 

Using the exact form of the equation of state\,\cite{Shapiro:1983du} the integral \eqref{logff} is immediate, and yields to leading order
\begin{align}\label{GBLfermions}
\frac{f(0)}{f(r_b)}\sim \frac{m ^2}{\sqrt{\rho(0)}}\,.
\end{align}

This is an important result, and has nothing to do with anomalies. For the Fermi gas, this is the essential relation defining the GBL. It shows how the center-to-surface redshift increases with the central density, eventually becoming infinite. 

For this equation of state, the crossover density \eqref{rhoc} is
\begin{align}\label{rhocfermi}
\rho_c\sim \left( m \, \Mp^2 \right)^{4/3} =  \left( \frac{m }{\Mp}\right)^{4/3} \rhop\,.
\end{align}
The small ratio $m /\Mp$ is what suppresses the Planck scale. Although this density is unimaginably higher than the neutron density, it is still much below the scale of quantum gravity. 

Therefore QFT in curved spacetime predicts that the curvature scalar becomes \textit{negative} in a neighbourhood of the origin at densities above \eqref{rhocfermi}. Until what density scales can we trust these results, e.g. \eqref{R0q}? The answer goes back to assumption (A3): we supposed that the individual components $T^{(A)}_{\mu\nu}$ are much smaller than the corresponding $T^{(m)}_{\mu\nu}$ of the fluid. This fails only at the Planck scale.

What are the physical consequences of this phenomenon? In the next section we will study the propagation of classical scalar waves on these backgrounds, and see that the anomaly induces a qualitative change in the behavior of their spectrum. 

%%%%%%%%%%%%

\section{Wave equation inside compact stars.}
\label{sec:wave}

The perturbation theory of relativistic stars is a rich and fascinating subject. At the classical level, it is well known that stars close to the Buchdahl limit are generically unstable. Here, we wish to explore the consequences of the quantum effect described above, where the trace anomaly modifies the behavior of the scalar curvature significantly.

As a first step towards a more general theory, we will examine the dynamics of a probe scalar field with no backreaction. We consider minimal coupling, the non-minimal generalization being straightforward\,\footnote{If we introduce a non-minimal coupling $\xi R$ in the wave equation, the conclusions are identical as the ones presented here for $\xi<1/6$. For $\xi>1/6$, they are 'reversed': e.g. unstable modes can appear above the crossover density, etc. If $\xi=1/6$, the trace anomaly plays no role at $r=0$. }. The wave equation
\begin{align}\label{wavescalar}
\Box \Phi=0
\end{align}
in the background of compact stars close to the GBL has remarkable properties\,\cite{1991RSPSA.434..449C}. To see this, it is convenient to perform the usual decomposition,
\begin{align}
\Phi=\sum f_{\omega\ell m}\,,\qquad f_{\omega\ell m}(x)=\frac{u(r)}rY_{\ell m}(\theta,\phi)e^{-i\omega t}
\end{align}
and to change variables to the `tortoise' coordinate $r_*$ via
\begin{align}
\frac{d r_*}{d r}=\sqrt{h(r)/f(r)} \,.
\end{align}

Then the wave equation takes the standard Schr\"{o}dinger form,
\begin{align}\label{Schr}
(-\partial_*^2+V_\ell)u=\omega^2 u
\end{align}
with the effective potential
\begin{align}\label{VtortoiseR}
V_\ell=\frac{f}{2rh}\left( \frac{f'}{f} - \frac{h'}{h} \right) + \frac{\ell(\ell+1)}{r^2}  f(r) \,.
\end{align}

The first terms, independent of the angular momentum, are smooth at the center due to the boundary conditions on the metric as discussed after \eqref{ds2static}. Outside the star, the potential is the usual one for scalars on a Schwarzschild vacuum metric.

\begin{figure}
      \centering
      \includegraphics[width=.8\columnwidth]{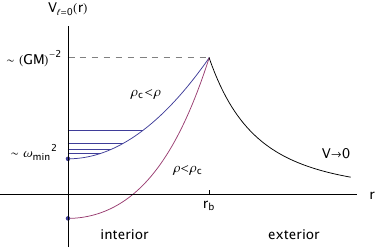}
      \caption{ Schematic representation of the radial profile of the potential in \eqref{VtortoiseR} for $\ell=0$. For $\rho \lesssim \rho_c$, the central potential is negative. The role of the anomaly for $\rho_c \lesssim \rho$ is to increase the central potential, so the quasi-bound state frequencies are bounded below by $\sim\omega_{\text{min}}$. }
      \label{fig:Vplot}
\end{figure}

For $\ell\neq 0$, the presence of the centrifugal barrier makes it difficult to evaluate $V_\ell$ without having the explicit solution for the metric. Therefore we focus on $\ell=0$. Fig. \ref{fig:Vplot} shows a schematic representation of the potential's profile for zero angular momentum. Evaluating at the center and imposing the regularity conditions \eqref{conditions} one finds
\begin{align}\label{V00}
V_{\ell=0}(0) = -\frac{1}{6} f(0) R(0)  \,.
\end{align}
Remarkably, the central potential is proportional to the curvature scalar, precisely the quantity determined by the anomaly!

To illustrate the physical implications of this, let us examine again the Fermi gas near the crossover density \eqref{rhoc}, and compare the behavior of the system with and without the conformal anomaly included into the field equations. In both cases, $f(0)$ is given at leading order by \eqref{GBLfermions} as we approach the GBL. Without the anomaly, we would find
\begin{align}\label{}
V_{\ell=0}(0)|_{a=0}\approx -\frac{1}{6}\left( \frac{m^2}{\Mp}\right)^2
\end{align}
at leading order in density. 

But once the anomaly becomes important, the central potential changes sign and grows positive instead due to the curvature scalar. This is illustrated in Fig. \ref{fig:Vplot}. As can be seen from the figure, the essential role of the anomaly is to make the gravitational potential `less attractive' to the scalar field. %Although this result only holds strictly for a small neighborhood around the origin and we lack the information about the anomaly throughout the rest of the star, from it we can deduce a number of interesting results regarding the spectrum of \eqref{Schr} localized around the origin. 

%The first result is the non-existence of unstable modes with $\omega^2<0$. For certain choices of boundary conditions for the scalar field, the wave equation \eqref{Schr} supports normalizable bound states localized around the origin, if the potential is negative enough. These grow exponentially with time and eventually backreact causing an instability. Since the anomaly dictates that the central potential eventually becomes positive, these instabilities cannot occur above the critical density. 

The anomaly controls the spectrum of resonances of \eqref{Schr}, the set of long-lived quasi-bound states of low frequencies, akin to their gravitational counterparts\,\cite{1991RSPSA.434..449C}.  The number of modes and their frequencies  -- measurable by an outside observer via spectroscopy -- change as the anomalous central potential grows. A simple estimate for the order of magnitude these frequencies is
\begin{align}\label{wmin}
\omega_{\text{min}}\sim\frac{m^2}{\Mp}\,,
\end{align}
because the potential vanishes at the crossover scale \eqref{rhoc}. As an example, $\omega_{\text{min}}\sim 10^{-2}s$ for the electron. So even though the energy scale of the processes occurring at the center of the star is enormous, an outside observer can perform low energy classical experiments to test this effect. This is directly related to the Generalized Buchdahl Limit: waves traveling from the origin outwards must climb an equally enormous gravitational potential to exit the star. Additional physical consequences of the effect described in this work are under investigation.

%The most interesting effect in this regard is when the potential at the center -- dominated by a quantum behavior -- equals the potential at the surface $r=r_b$. The latter knows only about classical gravity, $V_{\ell=0}(r_b)\sim (GM)^{-2}$ where $M$ is the mass of the star. Even for densities such that $\rho/\rho_c\gtrsim 1$ (and thus much below the Planck scale) this occurs when
%\begin{align}\label{Mgrav}
%M\sim \left( \frac{\Mp}{m} \right)^2 \Mp \,.
%\end{align}

%Although we ignore the QFT contributions throughout the rest of the star, this shows that the potential at the origin may become much larger than its value at the surface.

%If so, the effective potential experienced by the probe field in the quantum backreacted spacetime will resemble that of a much \textit{less compact} object in classical GR, even though the actual curvatures in the former are much higher than in the latter. This is an example where effects of QFT in curved spacetime can become of the same order as purely classical gravity ones. 

%In summary our results suggest that the anomaly modifies the geometry of a star in such a way as to make it more `repelling' than its non-anomalous version. %Indeed, the metric backreacted by the anomaly is less able to trap localized or quasi-localized states, particularly at low frequencies. % .... \textbf{GLUE} while \eqref{Mgrav} gives $M\sim 10^6 M_\odot$. Both scales seem relevant for astrophysical observations.

%\section{Acknowledgements}

%I am grateful to Julio Arrechea, Carlos Barcel\'o, Pablo Bosch, Ben Freivogel, Gerardo Garc\'ia-Moreno, Stefan Theisen, and the UvA string theory group for insightful discussions. In particular I wish to thank Gimmy Tomaselli, who participated in the initial stages of this project and helped to clarify many subtle points.   

\bibliography{Anomaly1_refs.bib}

\end{document}